\documentclass{elsart}
\usepackage{epsfig}
\usepackage{natbib}
\usepackage{amssymb}
\journal{Physica A}

\begin{document}

\begin{frontmatter}

\title{The co-evolutionary dynamics of directed network of spin market agents}

\author{Denis Horv\'ath \corauthref{cor}} 
\corauth[cor]{Corresponding author.}
\ead{horvath.denis@gmail.com}
\author{, Zolt\'an Kuscsik, Martin Gmitra} 
\address{ Dept. of Theoretical Physics and Astrophysics,          
 \v{S}af\'arik University, Park Angelinum 9, 040 01 Ko\v{s}ice, Slovak Republic}

\begin{abstract}
The spin market model~[S.~Bornholdt, Int.J.Mod.Phys. C {\bf 12} (2001) 667]
is extended into co-evolutionary version, where 
strategies of interacting and competitive traders 
are represented by local and global 
couplings between the nodes of dynamic directed stochastic network. 
The co-evolutionary principles are applied in the frame of 
Bak - Sneppen self-organized dynamics [P.~Bak, K.~Sneppen, 
Phys. Rev. Letter {\bf 71} (1993) 4083] that includes the processes 
of selection and extinction actuated by the local (node) fitness. 
The local fitness is related to orientation of spin agent 
with respect to instant magnetization. The stationary 
regime characterized by a fat tailed distribution of the log-price 
returns with index $\alpha\simeq 3.6$ (out of the L\'evy range) is identified 
numerically. The non-trivial consequence of the extremal dynamics is 
the partially power-law decay (an effective exponent varies 
between $-0.3$ and $-0.6$) of the autocorrelation function 
of volatility. Broad-scale network topology with node degree 
distribution characterized by the exponent $\gamma=1.8$ from 
the range of social networks is obtained. 
\end{abstract}


\end{frontmatter}

The Ising model is the simplest model of the solid state 
that is an example of the N-body problem. From the perspective 
of economics this example has a great importance because 
it demonstrates that a basic interaction between the spins (agents) 
can bring a non-trivial collective phenomena. The parallels 
between the fluctuations in the economic and magnetic systems affords 
an application of the spin models to the market statistics. 
This motivation has been outlined by Cont and Buchaud in~Ref.\cite{contK}. 
There is suggested that the equilibrium distribution of the 
"super-spins" can be used to model the stock price 
fluctuations. The "super-spin" model 
has been extended by the Chowdhury and Stauffer~Ref.\cite{ChowdhuryK} by switching 
on interactions between the lattice spins. Further works in this 
direction~Refs.\cite{bornholdtK,kaizoji2002,Takaishi2005} 
have putted emphasis onto linkage of the notion of strategy 
(as a way of an agent's thinking) with the spin degrees of freedom. 
The combination of these particulars can be made on the basis 
of the famous concept of {\em minority game} Ref.\cite{mingame97}. 
The spin variant of minority game assumes the competition of 
the ferromagnetic (short-range) and antiferromagnetic (global) 
couplings that cause an intermittent price 
dynamics of the stylised market. The salient points of 
the model~Ref.\cite{kaizoji2002} are reported in below.

The activity of {\it fundamentalists} and {\it interacting traders} 
has been imposed.  Each interacting trader is characterized by 
a regular lattice site position $i$ and by the corresponding 
spin variable $S^{(i)}(t)$ from $\{-1,1\}$. 
When $S^{(t)}(i)=1$, the agent tends to sell a unit amount of 
the stock whereas $S^{(t)}(i)=-1$ abbreviates the buy order during a given
period.  The variable $S^{(t)}(i)$ 
is updated 
by an asynchronous heat-bath dynamics 
\begin{eqnarray}
S^{(t+1)}(i) 
=\left\{ 
\begin{array}{lll}
1     &  \quad\mbox{\small with probability}   
& \quad P(h^{(t)}(i)) \,\,,
\\
-1    &  \quad\mbox{\small with probability}  
& \quad 1-P(h^{(t)}(i))\,\,,
\end{array}
\right.
\end{eqnarray}
where $h^{(t)}(i)$ 
denotes the local field and $P(h)$ is the sigmoid function 
\begin{equation}
P(h) =\frac{1}{1 + \exp(- 2 \beta h) }
\end{equation}
that depends on the inverse fictitious temperature $\beta$. The local 
field $h^{(t)}(i)$ reads
\begin{equation}
h^{(t)}(i)= J 
\sum_{j \in {\rm nn}(i)}  S^{(t)}(j) 
-  \kappa  S^{(t)}(i)\, |\, m^{(t)}\, |\,.
\label{paramBornKaiz}
\end{equation}
Here $J>0$ is the exchange coupling of nearest neighbors 
${\rm nn}(i)$ and $\kappa>0$ is the coupling of the local field 
with instant magnetization 
\begin{equation}
m^{(t)}=  \frac{1}{L}\sum_{i=1}^L S^{(t)}(i)\,,
\label{kaiz_magn}
\end{equation}
where $L$ is the number of the sites at some lattice. When $\beta J$ is bellow 
the critical temperature, the term $\kappa S^{(t)}(i) | m^{(t)} |$ suppresses the ferromagnetic 
spin phase. For sufficiently large $\kappa$ the frustration between 
the small (spin) and large (lattice) scales yields steady-state bubble-like regime.
The determination of mutual orientation of spin and minority is only possible 
with use of integral global information. Mathematically, the negative $S^{(t)}(i) m^{(t)}$ 
signalizes that agent $i$ belongs to the minority. The integral information also 
affects the price of given stock written in the form
\begin{equation}
p^{(t)} = p^{\ast} \, \exp \left( \lambda m^{(t)}\, \right)\,\,,
\label{priceaccrel}
\end{equation}
where $\lambda$ is the ratio of the number 
of fundamentalists to interacting traders. 
The formula can be interpteted as it follows: 
the predominance of buy orders implies $m^{(t)}>0$ 
which causes that $p^{(t)}$ falls above the fundamental 
price $p^{\ast}$. Evidently, the negative $m^{(t)}$ 
corresponds to under-valued stock.  Consequently, 
the change 
$m^{(t+1)}-m^{(t)}$ is associated with 
the logarithmic price return $\ln [ p^{(t+1)}/p^{(t)}]$ 
in further.

\section{Co-evolutionary dynamics on network of interacting spin agents}

In order to enhance the realism of the model, we have suggested several 
modifications. The primarily inspiration for this originates from the evolutionary 
formulation of minority game that allows agents to adapt strategy according 
to past experience~Ref.\cite{Johnson1999}. As a result, the co-evolutionary 
spin market model is proposed that imposes the self-organized 
formation of the stationary distributions of {\em strategic variables} 
constituted by the set of {\em inherent fields} 
and {\em couplings}.

       The static regular lattice geometry 
       represents perhaps one 
       of the most unrealistic aspects 
       of aforementioned spin model. 
       To overcome this problem 
       in our approach 
       we take advantage 
       of directed network (graph) 
       of labeled nodes 
       $\Gamma=\{1,2,\ldots, L\}$,
       where node $i\in \Gamma$ attaches  
       via $N^{\rm out}$  
       directed links to its 
       neighbors 
       $X_n(i) \in \Gamma$, 
       $n=1,2,\ldots, N^{\rm out}$, 
       i.e. the graph 
       is $N^{\rm out}$-regular.  
       The links mediate 
       non-symmetric 
       spin-exchange couplings 
       $J_n^{(t)}(i)$ 
       through which the 
       agents may exchange 
       the game-relevant information. 
       Two 
       outgoing links 
       $X_1(i)=1+(i) \mbox{mod} L$,
       $X_2(i)=1+(L+i-2) \mbox{mod} L$ 
       of node $i\in \Gamma$ 
       are fixed 
       to guarantee the network 
       connectedness 
       at any stage of its evolution.  
       The dynamics 
       of reconnections of 
       the links 
       $X_n(i)$, $n\geq 3$ 
       (defined below) then develops 
       on the background 
       of the 
       quenched bidirectional 
       loop ($n=1,2$). 
       In addition, 
       the dynamics 
       is constrained to forbid 
       the self-connections ($X_n(i)=i$) 
       as well as 
       the multiple 
       connections 
       [$X_{n_1}(i)=X_{n_2 \neq n_1}(i)$].
       An important information 
       about the network statistics 
       can be obtained 
       by sampling the distribution of the node degrees.  
       For our purposes here, 
       we redefine 
       {\em node degree} 
       $k^{(\rm in)}(j)= 
       \sum_{i=1}^L 
       \sum_{n=1}^{N^{\rm out}} 
       \delta_{j,X_n(i)}$ 
       as a number that accounts exceptionally 
       for the incoming links of the node $j$. 
   
With the above topology in mind, 
the spin interaction 
is introduced via modified local 
field 
\begin{equation}
h^{(t)}(i)  = 
\frac{1}{N^{\rm out}} 
\sum_{n=1}^{N^{\rm out}} 
J_n^{(t)}(i)\, S^{(t)} \left(X_n^{(t)}(i) \right)\,+ 
h_0^{(t)}(i)+ 
{\tilde \kappa}^{(t)}(i)\,
m^{(t)}\,.  
\label{kaizo1}
\end{equation}
We see that 
the strategic variable ${\tilde \kappa}^{(t)}(i)$ 
is reminiscent to the previously introduced global term. 
The inclusion of the node bias $h_0^{(t)}(i)$ closely 
resembles a random field model. It should be noted 
that preliminary simulations have confirmed 
qualitatively that the set of generated distributions 
is irrespective to variations in functional form of the local 
field. 

\begin{figure}
\begin{center}
\includegraphics[width = 9.0cm,angle=0]{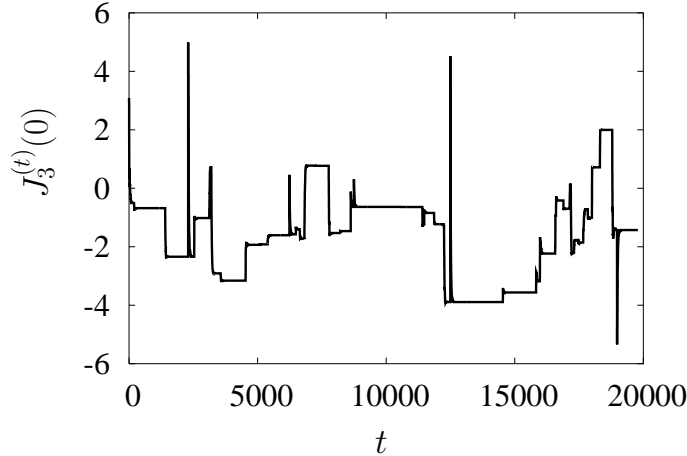}
\caption{ 
The varying coupling $J_3^{(t)}(0)$ shows punctuated 
equilibrium behaviour, i.e., periods of peaceful epochs 
interrupted by an extremely short fluctuating periods 
of boots and reconnections which are the hallmarks 
of the self-organized criticality. 
We inspected that the probability density function (pdf) of $J$'s indicates 
the formation of the "amorphous" and "glassy" mixture of the couplings 
that consists of ferromagnetic $(J_n^{(t)}(i)>0)$ and (prevailing) 
antiferromagnetic $(J_n^{(t)}(i)<0)$ contributions. 
The stationary mean value of $J^{(t)}_n(i)$ equal to $-0.518$ 
indicates the predominance 
of antiferromagnetic phase.}
\label{Fig1}
\end{center}
\end{figure}

\begin{figure}
\begin{center}
\includegraphics[width = 13.3cm,angle=0]{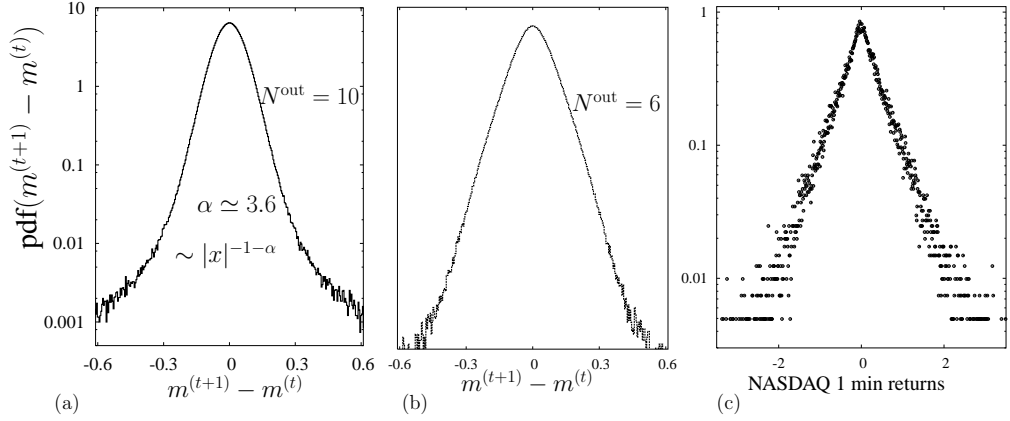}
\caption{The pdf of the log-price returns. 
In part (a) the index $\alpha=3.6$ of $N^{\rm out}=10$ 
links is extracted from $|x|^{-\alpha-1}$ fit. 
The central part of pdf is 
Gaussian distributed. Nearly exponential pdf in part 
(b) is obtained for $N^{\rm out}=6$. 
We see that pdf attained by pruning of links resembles 
qualitatively Nasdaq composite 
(c) of 1 min data through 
May - September 2005.}
\label{Fig2}
\end{center}
\end{figure}

\begin{figure}
\begin{center}
\includegraphics[width = 10.0cm,angle=0]{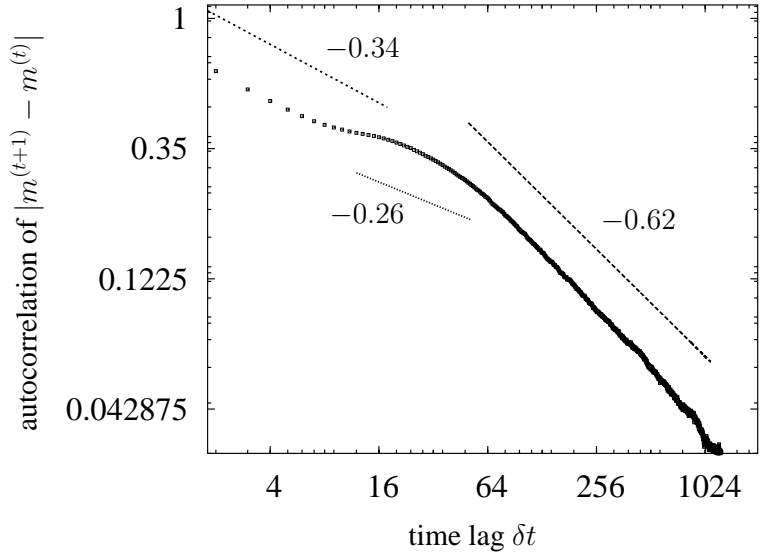} 
\caption{ The autocorrelation function of the volatility 
of the log-price returns $|m^{(t+1)}-m^{(t)}|$.  
The long-time memory with the particular 
effective decay exponents is identified 
numerically.} 
\label{Fig3}
\end{center}
\end{figure}

\begin{figure}
\begin{center}
\includegraphics[width = 9.1cm,angle=0]{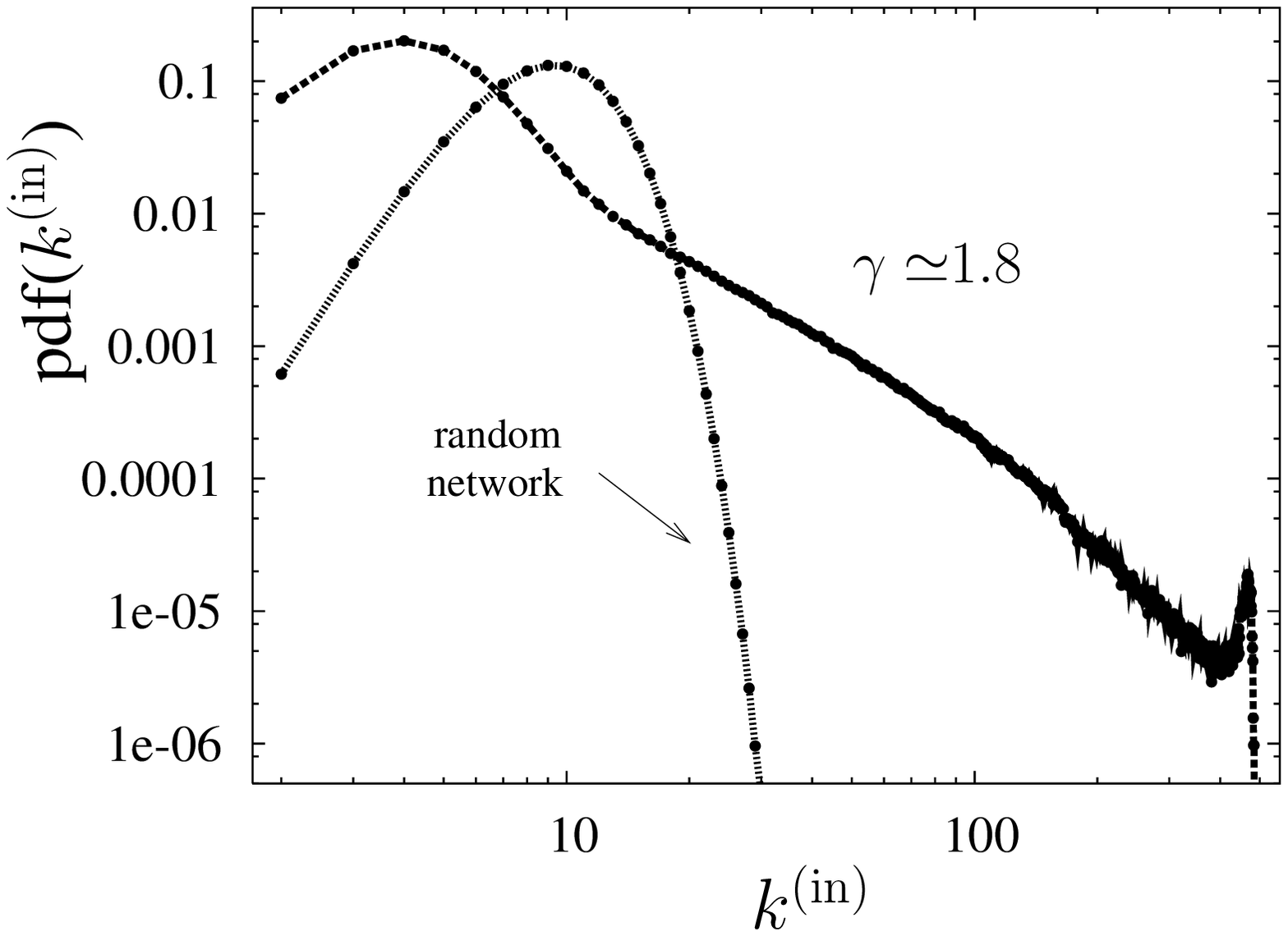} 
\caption{ 
The pdf's of the network node degree $k^{({\rm in})}(j)$ calculated for
$L=500$, $N^{\rm out}=10$. The comparison with the poissonian 
distribution obtained for the situation when the followers 
attach to randomly chosen node (no matter how large is the fitness). 
For the attachments to leader and its neighbors imposed the 
$[k^{({\rm in})}]^{-\gamma}$ tail is formed. The exponent 
$\gamma \sim 1.8$ corresponds to the fit 
within the range $<13,50>$.} 
\label{Fig5}
\end{center}
\end{figure}

As usual, the principles of the co-evolution are simply implementable in 
the frame of the local fitness as a phenotype. 
Here, the recurrently defined local fitness 
$F^{(t)}(i)$ is considered as an integrated history of
the agent's spin orientations with respect 
to the minority 
\begin{equation}
F^{(t+1)}(i) = F^{(t)}(i)  -  S^{(t)}(i)\, m^{(t)}\,.
\label{agent_fitness}
\end{equation}
The idea of the slowly (strategic) changing parameters~Ref.\cite{Krawiecki2002} 
as a background of the spin dynamics is not a new within 
the family of spin market models. In order to optimize 
the local fitness, the co-evolutionary dynamics including 
the Hebbian and anti-Hebbian learning rule, 
respectively, has been proposed in~Ref.\cite{Ponzi2000}. 
The approach proposed here follows similar 
lines, however, the principal 
distinctions exist that 
are summarized in below: 
\begin{description}

\item {\bf (a)} The strategies 
are subordinated to 
Bak - Sneppen extremal dy\-na\-mics 
Refs.\cite{Bak93,Flyvbjerg1993,Pienegonda2003};

\item {\bf (b)} In the present formulation we have $- S^{(t)} m^{(t)}$ 
instead of $- [\,S^{(t+1)}(i)-S^{(t)}(i)\,] m^{(t)}$  
suggested in~Ref.\cite{Ponzi2000}. 
The effect of the term involving the spin difference is that agent benefits from 
majo\-rity/mino\-rity jumps, while the abidance in minority is paradoxically 
penalized. 

\item {\bf (c)} The dynamical network 
of spin couplings is permanently 
rewired which mimics 
the social effects, where emphasis is putted on the links 
to the fittest nodes. Generally, our approach resembles the concept 
of preferential attachments that comes from 
the core 
of the proposal of the Barab\'asi and Albert~Ref.\cite{Albert99}.

\end{description}
 
Recently, the interest in complex networks has been extended 
to the search for local rules governing 
to dynamics of social networks. Several 
principles have been exploited for this purpose. 
As an example let us mention the network 
transformation due to node 
aging and time-dependent 
attachment probability~Ref.\cite{ZhuWangZhu2003}. 
The model proposed here is however closer 
to paradigm of inter-agent communication across 
net~Refs.\cite{Anghel2004,Zimmermann2004}. 

We have considered slow motion of {\em strategic variables} 
defined by the recursive cycles over the following steps:

\begin{enumerate}
 
\item  The information 
       that leader agent 
       labeled as 
       $i_{\rm max}$ 
       has achieved the highest 
       fitness 
       $F(i_{\rm max}) \geq F(j)$, $\forall j \in \Gamma$ 
       is available to all agents. 
       Clearly, the idealized and unbounded access 
       to information resources monitoring the instant 
       fitness of all agents is assumed. 

\item The 
      assumption of the model is that the strategic transfer 
      from {\em leader} 
      to the crowd of 
      $L-1$ followers 
      $i\in \Gamma$, $i\neq i_{\rm max}$ 
      is carried out independently 
      of the local network structure.
      The supposition about 
      the collective action of the crowd
      expresses the {\em bounded rationality} 
      of the {\em followers}. In essence, 
      all the followers 
      believe that imitation 
      of strategy that has been 
      applied in the aforegoing cycle 
      (i.e. after 1~visit per node) 
      by the 
      most successful 
      trader $i_{\rm max}$  
      would bring 
      a future benefit to them. 
      We postulated that each follower 
      $i$ adapts the strategy of 
      leader  
      according 
      to prescription    
\begin{eqnarray}
J_n^{(\rm next)}(i) &=& 
J_n(i) 
\left(\, 1- \eta_{\mbox{\tiny\mbox{$J,n$}}}(i) \,\right)  +   
J_n(i_{\rm max})  
\eta_{\mbox{\tiny \mbox{$J,n$}}}(i)\,,
\\
h_0^{(\rm next)}(i) &=&  
h_0(i)  
\left(  1 - 
\eta_{\mbox{\tiny\mbox{$h_0$}}}(i)\,  \right)  
+  h_0(i_{\rm max}) 
\eta_{\mbox{\tiny\mbox{$h_0$}}}(i)\,,
\\
{\tilde \kappa}^{(\rm next)}(i) &=&  
{\tilde \kappa}(i) 
\left(\,  
1- \eta_{\mbox{\tiny\mbox{$\tilde\kappa$}}} (i)  \, 
\right)  
+ {\tilde \kappa}(i_{\rm max}) 
\eta_{\mbox{\tiny\mbox{$\tilde\kappa$}}}(i)\,,
\\
P_{\rm lead}^{(\rm next)}(i) &=& P_{\rm lead}(i) 
\left(\,  1- \eta_{\mbox{\tiny \mbox{$PL$}}} (i) \,\right)  
+ 
P_{\rm lead}(i_{\rm max}) \eta_{\mbox{\tiny\mbox{$PL$}}}(i)\,,
\\
P^{(\rm next)}_{\rm T}(i)  &=& 
P_{\rm T}(i) 
\left(\,  1- \eta_{\mbox{\tiny \mbox{$PT$}}} (i) \,\right)  
+   
P_{\rm T}(i_{\rm max}) 
\eta_{\mbox{\tiny\mbox{$PT$}}}(i)\,,
\label{Cross1}
\end{eqnarray}
The formula 
expresses the individual adaption differences. 
They are reflected by the plasticity parameters 
$\eta_{\mbox{\tiny\mbox{$J,n$}}}(i)$,  
$\eta_{\mbox{\tiny\mbox{$h_0$}}}(i)$, 
$\eta_{\mbox{\tiny\mbox{$\tilde\kappa$}}}(i)$
$\eta_{\mbox{\tiny\mbox{$PT$}}}(i)$,
$\eta_{\mbox{\tiny\mbox{$PL$}}}(i)$
uniformly distributed over the range $(0,\eta_{\rm max})$, where 
$0<\eta_{\rm max}<1$. 

\item What have been left in the previous step 
      without explanation 
      are local probabilities 
      of preferential attachments 
      $P_{\rm lead}(i)$, 
      $P_{\rm T}(i)$. 
      They occur in the  
      specific local rules 
      of network dynamics. 
      For given $i$  
      and randomly 
      selected $n, n_{\rm 1}\geq 3$  
      the rewiring 
      follows 
      one of  
      three 
      probabilistic 
      channels
\begin{eqnarray}
\begin{array}{llll}
X_n(i)  &   \mbox{{\scriptsize 
                  \mbox{ $\left[ 
                       \begin{array}{l} 
                           \mbox{ \scriptsize is left }
                            \\
                            \mbox{ \scriptsize without change} 
                             \\
                             \end{array} \right] $ } }}  & 
                 \mbox{\small with probability} & 
                 1 - P_{\rm lead}(i)\,,
\\
X_n(i) & \leftarrow i_{\rm max}   & 
          \mbox{\small with probability}   &   (1- P_{\rm T}(i) ) \, P_{\rm lead}(i)\,,
\\
X_{n_1\neq n}(i)  &  \leftarrow  X_{n_1}(i_{\rm max})  & \mbox{\small with probability} & P_{\rm lead}(i)\, P_{\rm T}(i)\,,
\\
\end{array}
\label{rulenetw1}
\end{eqnarray}
where $(1-P_{\rm T}(i)) P_{\rm lead}(i)$ 
is the probability of the connection to node of actual leader. 
The update $X_n(i) \leftarrow X_{n_1}(i_{\rm max})$ 
favors the {\em social transitivity}~Ref.\cite{Ebel2003}.  
In addition, the transitivity requires bounding 
$P_{\rm T}(i)  \leq \epsilon_{\rm T} \sim  O(1) \gg P_{\rm lead}$.
 
\item At given $t$ the synchronous update 
$J_n(i) \leftarrow J_n^{(\rm next)}(i)$,  
${\tilde\kappa}(i) \leftarrow {\tilde \kappa}^{(\rm next)}(i)$, 
$ h_0(i) \leftarrow  h_0^{(\rm next)}(i)$, 
$ P_{\rm lead}(i) \leftarrow  P_{\rm lead}^{(\rm next)}(i)$,
$ P_{\rm T}(i) \leftarrow  P_{\rm T}^{(\rm next)}(i)$
is carried out for all $i\in 
\Gamma$ and $n=1,2,\ldots, N^{\rm out}$. 

\item The agent $i_{\rm min}$ 
of the lowest fitness $F(i_{\rm min})$ finishes her/his 
unsuccessful life via 
{\em extremal dynamics}.  
The death-birth rules 
$ J^{(t+1)}_n(i_{\rm min}^{(t)}) 
= ( 2 r_{1,n}^{(t)} - 1 ) \epsilon_{\rm J} $, 
$ h_0^{(t+1)}(i_{\rm min}^{(t)})  = ( 2 r_3^{(t)} - 1 ) \epsilon_{{\rm h}_0} $, 
$ {\tilde \kappa}^{(t+1)}(i_{\rm min}^{(t)})  =  ( 2 r_4^{(t)}-1  ) \epsilon_{\kappa} $
with integer $X^{(t+1)}_n(i_{\rm min}^{(t)}) = $ 
$ [\, 1+L r_{2,n}^{(t)} ]_{\rm int}$ and 
strictly positive 
$P_{\rm lead}^{(t+1)}(i_{\rm min}^{(t)}) = 
r_5 \epsilon_{\rm lead}$, 
$P_{\rm T}^{(t+1)}(i_{\rm min}^{(t)}) = 
r_6^{(t)} \epsilon_{\rm T}$
are driven by the uncorrelated uniformly 
distributed numbers 
$\{  r_{1,n}^{(t)}  \}_{n=1}^{N^{\rm out}}$, 
$\{  r_{2,n}^{(t)}  \}_{n=1}^{N^{\rm out}}$, 
$\{ r_j^{(t)}\}_{j=3}^{7}$ from $(0,1)$. 
The last $r_7^{(t)}$ serves for "nearly middle" setting 
$F^{(t+1)}(i_{\rm min}^{(t)})= F^{(t)} ( i_{\rm min}^{(t)}) + 
r_7^{(t)} ( F(i_{\rm max}^{(t)})- F(i_{\rm min}^{(t)})  )$.
The random numbers depend on the free parameters 
$\epsilon_{\rm J}, \epsilon_{{\rm h}0}, 
\epsilon_{\kappa}, \epsilon_{\rm lead}, \epsilon_{\rm T}$ 
that bound the available strategic space. 
The same updates have been used for imposition 
of initial conditions. As one can see from the structure of updates, 
the adaptivity in common with extremal 
dynamics preserves 
the span of the strategic space. 

\end{enumerate}

\section{Numerical results}

The simulation has been carried out for the spin system thermalized 
via the asynchronous Glauber dynamics with $\beta=1$
within the strategic bounds $\epsilon_{\rm lead}=0.01$, 
$\epsilon_{\rm T}=0.5$, $\epsilon_{\rm J}=8$, 
$\epsilon_{\kappa}=\epsilon_{{\rm h}_0}=4$, and with topology $L=500$, $N^{\rm out}=10$ 
for $t \leq 2 \times 10^7$ iterations
initialized from random initial settings [see step (5) of the algorithm].   
As usual, the data corresponding to transient regime has been discarded. 
Hereafter we summarize the main findings of simulation.

       The self-organization 
       process of intermittent
       fluctuations 
       of the log-price returns 
       has been obtained by the simulation.  
       The punctuated equilibrium 
       of selected strategic 
       variable is monitored in Fig \ref{Fig1}. 

       The relevance 
       of the approach for 
       the economic modeling is demonstrated 
       by the fat tail probability density function (pdf) 
       of the log-price returns~Ref.\cite{Mantegna95}.  
       It can be roughly characterized by the exponent $\alpha\simeq 3.6$ (Fig.\ref{Fig2}). 
       At this moment, it is rather illustrative to mention the empirical 
       value $\alpha=3.1$ corresponding to 1-day period 
       of TAQ database~Ref.\cite{Stanley2000}.
       It has been checked numerically 
       that distribution may vary with the 
       bounding of strategic space 
       and chosen topology ($N^{\rm out}$).
       The generalization proposed herein covers not only power-law 
       but also nearly exponential or Gaussian-like distributions.
       
       In Fig.\ref{Fig3} the emergence of the long market memory 
       is demonstrated by calculating the autocorrelation 
       of volatility of the log-price returns. The empirical slopes 
       $-0.34$~Ref.\cite{Stanley2000}
       and $-0.2$~Ref.\cite{Zawadowski2004} should be mentioned 
       in this context. As it turns out from our simulation, 
       the co-evolution represents 
       one of the potential mechanisms responsible for the 
       power-law decay. More specifically, we have shown   
       the inclusion of 
       {\em extremal dynamics and adaptivity} 
       can cure 
       the notorious exponential decay problem 
       of the spin market models.  
       Even more interestingly, only 
       the {\em simultaneous exploiting of both mentioned 
       dynamical principles can result in 
       nontrivial autocorrelations}.
       Within the spin market models, 
       an alternative explanation 
       to our has been proposed 
       only recently~Ref.\cite{Takaishi2005}. 
       In the quoted work, 
       the power-law decay is kept
       for three-state Potts model 
       where zero spin state encodes 
       the inactive (waiting) 
       state of the trader.
  
       In the stationary regime the self-organized reconstruction 
       yields network of the broad-scale distributed  
       degrees of nodes~(see Fig.\ref{Fig5}). For the upper part of the power-law range 
       we have estimated an exponent $\gamma\simeq 1.8$. The result is quite 
       encouraging since the exponent falls within the range typical 
       for the social networks. Here we report on a comparison the exponents 
       of several cooperative social networks:
       the value $\gamma=1.81$ Ref.~\cite{Ebel2002} 
       is obtained by collecting the e-mal addresses, 
       the exponent $1.2$ belongs to the 
       coauthorship 
       network Ref.~\cite{Newman2001} 
       whereas $2.1$ characterizes 
       the phone-call 
       network 
       Ref.~\cite{Aiello2000}.
       Similarly 
       as in~Ref.\cite{DorogMendes04}, 
       the network topology 
       can be characterized 
       by the clustering coefficient $C_i$ 
       or its spatial-temporal mean 
       $\langle C \rangle $. 
       For directed network 
       we have used 
       slightly modified formula   
       $C_i = e_i/ ( N^{\rm out} (1-N^{\rm out})) $,  
       where 
       $e_i  =  \sum_{n_1,n_2,n_3=1}^{N^{\rm out} \times  
       N^{\rm out} \times N^{\rm out}} 
       \delta_{X_{n_1}(X_{n_2}(i)),X_{n_3}(i)}$ 
       stands for the number of 
       links between the neighbors 
       of some node $i$. 
       In this formula, 
       $N^{\rm out} \,( 1 - N^{\rm out} )$ 
       represents the maximum number of links 
       going from $X_{n_2}(i)$ to 
       $X_{n_3}(i)$ and vice versa. 
       It should be noted that the conventional 
       factor $2$ absents in the definition 
       since the 
       links are directed. As usual, it is meaningful 
       to compare the mean clustering coefficients  
       of two simulated network regimes. 
       For randomly attached 
       nodes [instead of selection declared by Eq.(\ref{rulenetw1})]
       the simulation results in  $\langle C_{\rm rand} \rangle \simeq
       0.027$, 
       while 
       $\langle C \rangle \simeq 0.125$ 
       stems from 
       the preference of 
       $i_{\rm max}$, and thus the enhancement of clustering  
       is confirmed by
       the ratio 
       $\langle C \rangle /\langle C_{\rm rand} \rangle \simeq 4.6$.

\section{Conclusions}

The generalized version of the spin market model has been suggested and investigated numerically. 
In the approach we used the static couplings are adjusted by an instant 
local fitness that drives the co-evolutionary competitive changes. The socially 
relevant feature of the approach is slowly evolving topology of the spin-spin interaction network 
as a substrate for fast buy-sell spin decisions. The power-law distributions have 
been uncovered as an attributes of the universality caused by the extremal dynamics. 
Our results suggest hypothesis that slowly varying strategic variables can cause 
an emergence of the long memory effect that is symptomatic for the real markets. 
The universality manifests itself in a broad-scale node degree distribution, however, 
{\em the network evolution does not necessary imply the power-law decay of volatility}. 
Even the static net enough to gain the power-law decay or flat-tailed pdf's. 
Numerous examples of the model applicability can be found outside the field of econophysics. 
For instance, the coevolutionary model opens the possibility of the optimization 
of the spin lattice and related systems in the manner of~Ref.\cite{Onody2003}. 
The finite-size scaling analysis, the small world behaviour and the aspect
of modularity are only a few examples of issues that remain to be analyzed in 
the complementary studies. 

\ack
The authors would like to express their thanks to Slovak Grant agency
VEGA (grant no.1/2009/05) and agency APVT-51-052702 for financial support.


\begin{thebibliography}{99}

\bibitem{contK} R.~Cont, J.P.~Bouchaud, preprint cond-mat/9712318; page 71 in 
J~.Bouchaud and M.~Potters Theories des Risques Financiers (Alea Saclay/ Eyrolles, 1997).

\bibitem{ChowdhuryK} D.~Chowdhury, D.~Stauffer, European Physical Journal B {\bf 8} (1999) 477. 

\bibitem{bornholdtK} S.~Bornholdt, Int.J.Mod.Phys. C {\bf 12} (2001) 667.

\bibitem{kaizoji2002} T.~Kaizoji, S.~Bornholdt, Y.~Fujiwara,  
                      Physica A {\bf 316} (2002) 441.

\bibitem{Takaishi2005} T.~Takaishi, cond-mat/0503156.

\bibitem{mingame97} D.~Challet, 
                    Y.-C.~Zhang, Physica A {\bf 246} (1997) 407; 
                    D.~Challet, M.~Marsili, Phys. Rev. E {\bf 60} (1999) R 6271.

\bibitem{Johnson1999} N.F.~Johnson, P.M.~Hui., R.~Jonson, 
                      T.S.~Lo, Phys. Rev. Lett {\bf 82} (1999) 3360.

\bibitem{Krawiecki2002} A.~Krawiecki, 
                        J.A.~Holyst, 
                        D.~Helbing,  
                        Phys. Rev. Lett. 
                        {\bf 89} (2002) 158701.

\bibitem{Ponzi2000} A.~Ponzi, Y.~Aizawa,  Physica A {\bf 287} (2000) 507.

\bibitem{Bak93} P.~Bak, K.~Sneppen, Phys. Rev. Lett. {\bf 71} (1993) 4083.

\bibitem{Flyvbjerg1993} H.~Flyvbjerg, K.~Sneppen, P.~Bak Phys. Rev. Lett. {\bf 71} (1993) 4087.

\bibitem{Pienegonda2003} S.~Pienegonda, J.R.~Iglesias, 
                         G.~Abramsovic, J.L.~Vega, 
                         Physica A {\bf 393} (2003) 667. 

\bibitem{Albert99} A.L.~Barab\'asi, R.~Albert, Science {\bf 286} (1999) 509.

\bibitem{ZhuWangZhu2003} H.~Zhu, X.~Wang, J.Y.~Zhu, Phys. Rev. E {\bf 68} (2003) 056121. 

\bibitem{Anghel2004} M.~Anghel, Z.Toroczkai, 
                     K.E.~Bassler, G.~Korniss, Phys. Rev. Lett {\bf 92} (2004) 058701.

\bibitem{Zimmermann2004} M.G.~Zimmermann, 
                         V.M.~Egu\'{\i}luz, M.S.~Miguel, Phys. Rev. E {\bf 69} (2004) 065102(R).

\bibitem{Ebel2003} H.~Ebel, J.~Davidsen, S.~Bornholdt, Complexity {\bf 8} (2003) 24.

\bibitem{Mantegna95} R.N.~Mantegna, H.E.~Stanley, Nature {\bf 376} (1995) 46.

\bibitem{Stanley2000} H.E.~Stanley, P.~Gopikrishnan, V.~Plerou, L.A.N.~Amaral, Physica A {\bf 287}, (2000) 339.

\bibitem{Zawadowski2004} A.G.~Zawadowski, J.~Kert\'esz, G.~Andor, Physica A {\bf 344} (2004) 221.

\bibitem{Ebel2002}  H.~Ebel, L.I.~Mielsch, and S.~Bornholdt, Phys. Rev. E {\bf 66}, (2002) 035103. 

\bibitem{Newman2001} M.E.J.~Newman, Phys.Rev.E {\bf 64} (2001) 016131.

\bibitem{Aiello2000} W.~Aiello, F.R.K.~Chung, L.~Lu, in {\it 32}nd 
                     {\it ACM} 
                     {\it Symposium on Theory of Computing} 
                     (ACM, New York, 2000), 
                     pp.171-180.

\bibitem{DorogMendes04} S.N.~Dorogovtsev, J.F.F.~Mendez, cond-mat/0404593.

\bibitem{Onody2003} R.N.~Onody, P.A.~Castro, Int. J. Mod. Phys. C {\bf 14} (2003) 911.  



\end{thebibliography}
\end{document}